\documentclass[US Letter]{article}
\usepackage{spconf,amsmath,graphicx,amssymb,delarray,verbatim,booktabs}
\usepackage{diagbox, makecell}
\usepackage{multirow}
\usepackage[english]{babel}
\usepackage{lipsum}

\title{TWO HEADS ARE BETTER THAN ONE: A TWO-STAGE APPROACH FOR MONAURAL NOISE REDUCTION IN THE COMPLEX DOMAIN}
\name{Andong Li$^{\star \dagger}$, Chengshi Zheng$^{\star \dagger}$, Renhua Peng$^{\star \dagger}$, Xiaodong Li$^{\star \dagger}$}
\address{$^{\star}$ Key Laboratory of Noise and Vibration Research, Institute of Acoustics, Chinese Academy\\
	of Sciences, Beijing, China\\
	$^{\dagger}$ University of Chinese Academy of Sciences, Beijing, China}
\begin{document}
\ninept

\maketitle
\newcommand\blfootnote[1]{%
	\begingroup
	\renewcommand\thefootnote{}\footnote{#1}%
	\addtocounter{footnote}{-1}%
	\endgroup
}

\begin{abstract}
In low signal-to-noise ratio conditions, it is difficult to effectively recover the magnitude and phase information simultaneously. To address this problem, this paper proposes a two-stage algorithm to decouple the joint optimization problem $\emph{w.r.t.}$ magnitude and phase into two sub-tasks. In the first stage, only magnitude is optimized, which incorporates noisy phase to obtain a coarse complex clean speech spectrum estimation. In the second stage, both the magnitude and phase components are refined. The experiments are conducted on the WSJ0-SI84 corpus, and the results show that the proposed approach significantly outperforms previous baselines in terms of PESQ, ESTOI, and SDR.
\end{abstract}
\vspace{-0.2cm}
\section{Introduction}
\label{sec:intro}
\vspace{-0.2cm}
Various interference such as environmental noise and room reverberation may result in heavy performance degradation in automatic speech recognition (ASR) and hearing assistance devices. Despite endless efforts in the last five decades, speech enhancement (SE), a type of technique to extract the clean version from the noisy speech, still remains a tough challenge under low signal-to-noise ratio (SNR) and nonstationary noise conditions~{\cite{loizou2013speech}}.

In recent years, the rapid development of deep neural networks (DNNs) has facilitated the research toward supervised speech enhancement algorithms~{\cite{wang2018supervised, kolbaek2020loss, xu2014regression}}. In the conventional supervised SE paradigm, DNN is leveraged to extract a clean feature from a noisy observation in the time-frequency (T-F) domain, $\emph{e.g.}$, ideal ratio mask (IRM)~{\cite{hummersone2014ideal}} or log-power spectrum (LPS)~{\cite{xu2014regression}}. With this aim, these methods usually focus on the estimation of the spectrum magnitude, and the noisy phase is left unaltered to reconstruct the waveform in the time domain.

More recently, the importance of phase begins to be emphasized as it is helpful to improve the perceptual quality under low SNR conditions~{\cite{paliwal2011importance}}. However, phase is usually difficult to estimate due to its unstructured characteristics. To this end, some SE algorithms are thus proposed and can be roughly divided into two categories. The first one tries to recover the phase information implicitly in the frequency domain, where both real and imaginary (RI) components of the spectrum are taken into optimization when using the complex spectral mapping based networks~{\cite{tan2019learning}}. The second one is to divert around direct phase estimation problem in the time domain, where the time-domain waveform is utilized as both the input and the output~{\cite{kolbaek2020loss, pandey2020densely}}. It has shown that noise and speech components tend to be more distinguishable in the T-F domain, and meanwhile, it is also easier for network to train in the T-F domain~{\cite{yin2020phasen}}, and thus this study focuses on SE in the T-F domain.



Multi-stage learning has been demonstrated to be more effective than single-stage methods in many tasks, like image deraining~{\cite{li2018recurrent}} and SE~{\cite{li2020speech, li2020recursive, hao2020masking, 9043689}}. In multi-stage learning protocol, the original difficult task is decomposed into multiple easier sub-tasks and the estimated target is progressively improved. Motivated by that, we propose a \textbf{C}omplex spectral mapping based \textbf{T}wo-\textbf{S}tage \textbf{N}etwork called CTS-Net for monaural speech enhancement. It is comprised of two sub-networks, namely \textit{coarse magnitude estimation network} (dubbed CME-Net) and \textit{complex spectrum refine network} (dubbed CSR-Net). In the first stage, the target spectral magnitude (TMS) is coarsely estimated by CME-Net, which is then coupled with noisy phase to obtain a coarsely estimated complex spectrum (dubbed coarse spectrum). In the second stage, the complex spectrum is further refined by CSR-Net, which recovers both the real and imaginary (RI) components. Note that, instead of explicitly estimating the complex spectrum, we only estimate the residual details in the second stage. This is because most noise interference has been removed in CME-Net and the second network aims to recover the clean speech phase, further suppress the residual noise, and restore some missing spectral details in the first stage.

The rationale behind such network design logic can be explained from two aspects. On one hand, it has been illustrated that an optimal solution for multi-task $\emph{w.r.t.}$ magnitude and phase can not be obtained simultaneously~{\cite{wang2020complex}}, especially in extremely low SNR condtions. For example, while the phase is continuously optimized by estimating RI components, magnitude estimation may gradually deviate its optimal optimization path. Therefore, the two-stage network topology can decouple the multi-task optimization problem w.r.t. magnitude and phase into two sub-tasks. On the other hand, it has been revealed that there exists a latent connection between magnitude and phase in the T-F domain~{\cite{griffin1984signal}}, and thus this relatively ``clean" magnitude obtained by the first stage can facilitate the phase recovery in the second stage.

The remainder of the paper is organized as follows. Section~{\ref{sec:problem-formulation}} formulates the problem. In Section~{\ref{sec:proposed-architecture}}, the proposed architecture is introduced in detail. Section~{\ref{sec:experiment-result}} gives the experimental results and analysis. Some conclusions are drawn in Section~{\ref{sec:conclusion}}.

\vspace{-0.4cm}
\section{PROBLEM FORMULATION}
\label{sec:problem-formulation}
\vspace{-0.2cm}
In the time domain, a single-microphone mixture $x$ is usually formulated as $x\left( t \right) = s\left( t \right) + n \left( t \right)$, where $s$ and $n$ denote clean and noise signal in the time index $t$. Taking the short-time Fourier transform (STFT) on both sides, we have:
\begin{equation}
\label{eqn:equa1}
X_{m, l} = S_{m, l} + N_{m, l},
\end{equation}
where $X_{m, l} = \lvert X_{m, l} \rvert e^{j\theta_{X_{m, l}}}$, $S_{m, l} = \lvert S_{m, l} \rvert e^{j\theta_{S_{m, l}}}$ and $N = \lvert N_{m, l} \rvert e^{j\theta_{N_{m, l}}}$ refer to the STFT representations of noisy, clean and noise, resepctively. $m$ and $l$ denote the frequency index and the time/frame index, respectively. For notation simplicity, we drop $\left(m, l\right)$ when no confusion arises.

Phase is usually left unchanged due to its unstructured characteristic for conventional SE algorithms. However, in Cartesian coordinates, phase can be implicitly represented by RI components, $\emph{e.g.}$, $\theta_{S} = \arctan \left(S_{i}/ S_{r}\right)$, which provides a promising direction for phase estimation. Recently, Tan $\emph{et.al.}$~{\cite{tan2019learning}} proposed a complex spectral mapping method with convolutional recurrent network (CRN), which took noisy and clean RI as inputs and targets, respectively. Assuming the network mapping function and its parameter set are $\mathcal{G}_{C}$ and $\phi_{C}$, respectively, the mapping process can be given by:
\begin{gather}
\label{eqn:equa2}
\left( \tilde{S}_{r}, \tilde{S}_{i} \right) = \mathcal{G}_{C}\left(X_{r}, X_{i};\phi_{C} \right),
\end{gather}
where $\left( \tilde{S}_{r}, \tilde{S}_{i} \right)$ denote the estimated RI.

\vspace{-0.2cm}
\section{PROPOSED ARCHITECTURE}
\label{sec:proposed-architecture}

\begin{figure}[t]
	\centering
	\centerline{\includegraphics[width=0.85\columnwidth]{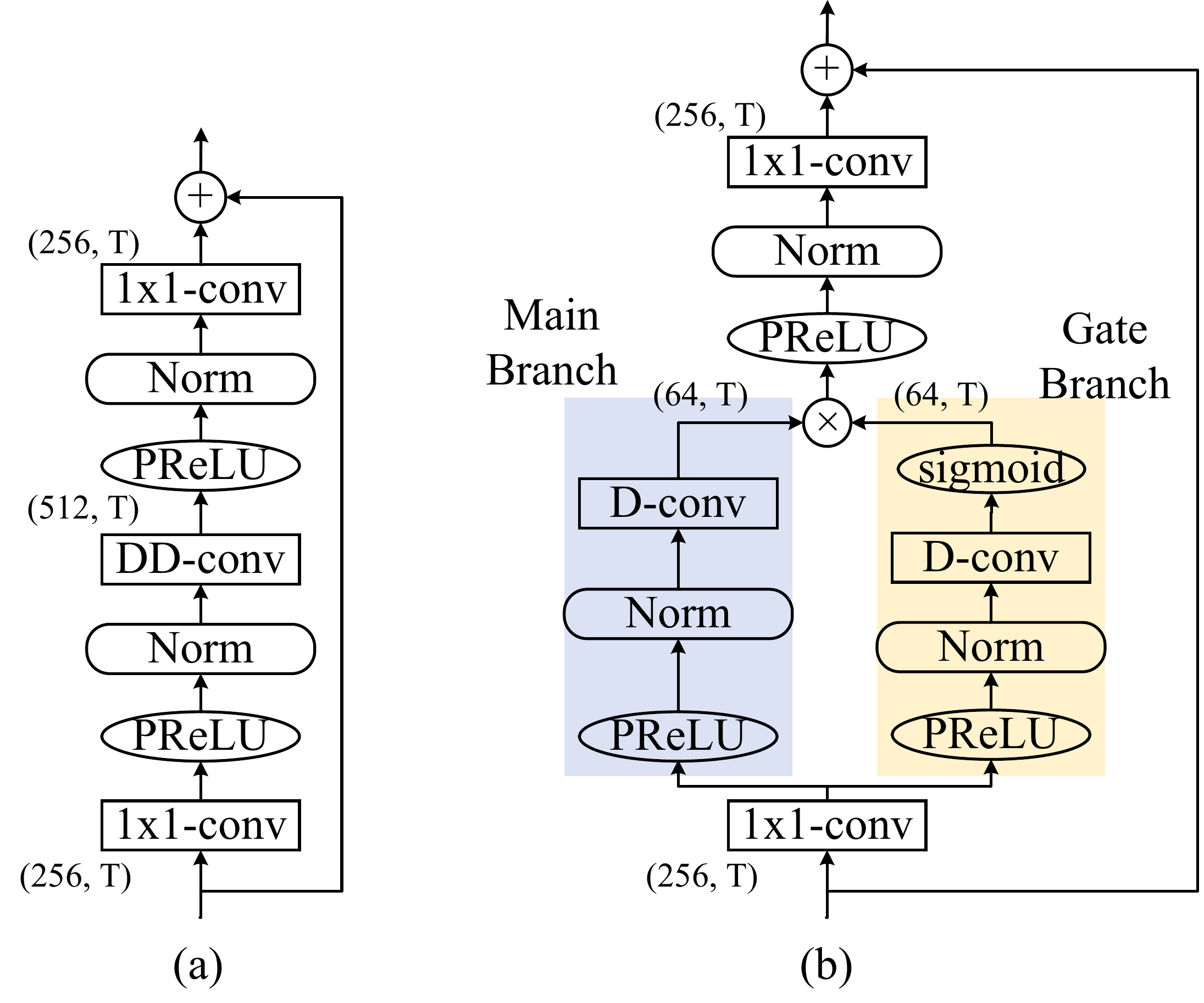}}
	\caption{Comparison between original TCM and proposed MG-TCM. (a) The diagram of original TCM. (b) The diagram of proposed MG-TCM.}
	\label{fig:gate-tcm}
	\vspace{-0.4cm}
\end{figure}
\vspace{-0.2cm}
\subsection{Modified gated temporal convolution module}
\label{sec:modified-gated}
\vspace{-0.0cm}
In~{\cite{luo2019conv}}, a convolution-based module named temporal convolution module (TCM) was proposed to replace traditional long-short term memory units (LSTMs) for better temporal sequence modeling. The diagram of TCM is shown in Fig.~{\ref{fig:gate-tcm}} (a). It consists of three major parts, namely input 1$\times$1-conv, dilated depthwise convolution (DD-conv) with kernel size 3, and output 1$\times$1-conv. Assuming the input size is $\left(256, T \right)$, where 256 and $T$ denote the channel and time axis, respectively. The number of channels is first doubled to 512, followled by a DD-conv. Then another 1$\times$1-conv layer is utilized to switch the channel back to 256. Residual connection is adopted to mitigate the gradient vanishing problem.

Despite the excellent performance of TCM, its drawback in heavy number of parameters is still noticeable. Specifically, although depthwise convolution is adopted to increase the parameter efficiency, as the channel dimension is increased to 512, the number of parameters in both input and output 1$\times$1-convs are still relatively large. To further ease the parameter burden, we propose a light-weight TCM variant named modified gated TCM (MG-TCM), as shown in Fig.~{\ref{fig:gate-tcm}} (b). Compared with original TCM (dubbed O-TCM), it has several modifications. Firstly, the channel dimension is decreased to 64 after the input 1$\times$1-conv, which notably decreases the parameter redundancy. Secondly, considering the limited ability for depthwise convolution in feature transform, we replace it with regular dilated convolution (D-conv) of kernel size 5 to grasp more information between adjacent frames. Thirdly, apart from the main branch, we also add another branch called gate branch, which has a similar structure as the main branch, except the sigmoid function is applied as the non-linearity to modulate the feature distribution. As shown in Fig.~{\ref{fig:gate-tcm}}, if neglecting the parameters for norm and activation layers, the number of parameters for O-TCM is 263,680, while only 53,248 for MG-TCM.

\begin{figure}[t]
	\centering
	\centerline{\includegraphics[width=0.85\columnwidth]{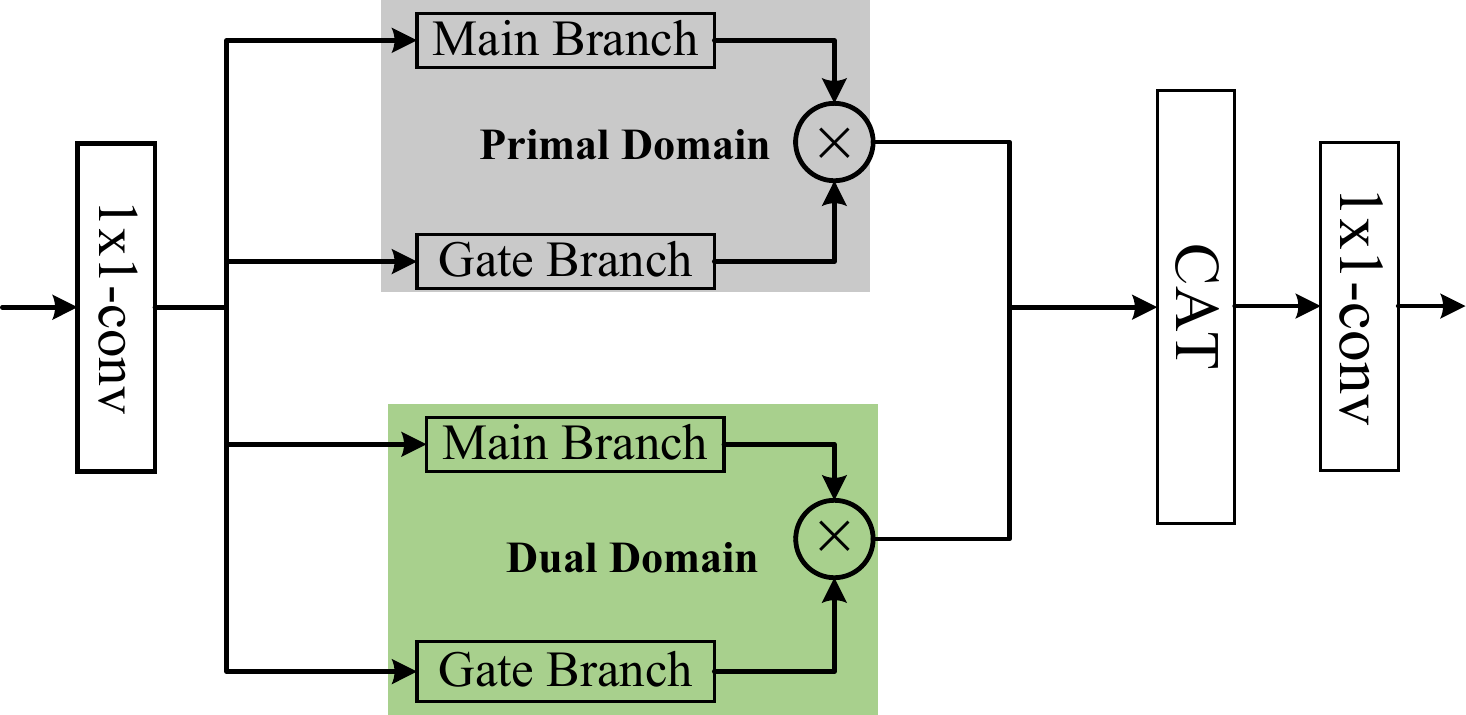}}
	\caption{Diagram of DMG-TCM. Norm and activation layers are omitted for convenience. ``CAT" denotes concatenation operation.}
	\label{fig:dmg-tcm}
	\vspace{-0.6cm}
\end{figure}

TCMs are usually stacked to grasp a larger temporal contextual field, where the dilation rate in each block is exponentially increased. If the dilation rate gets larger, then more attention is placed to long-term dependency, thus neglecting the correlation in the local regions. In this case, we further propose a dual version of MG-TCM named DMG-TCM to address this issue, as shown in Fig.~{\ref{fig:dmg-tcm}}. Compared with MG-TCM, it has two domains, namely primal and dual domain. The module detail within each domain is similar to that of MG-TCM except for the dilation rate. Suppose the dilation rate in the primal domain is $d_{\mathcal{P}} = 2^{r}$, then the dilation rate in the dual domain becomes $d_{\mathcal{D}} = 2^{M-r}$. In our method, $M = 5$ suffices empirically. When $d_{\mathcal{P}}$ in primal domain is small, $\emph{e.g.}$, $d_{\mathcal{P}} = 1$, the sequence in primal domain will pay more attention to local correlation, then $d_{\mathcal{D}}$ in dual domain becomes $32$, so more information among long-temporal correlation can be grasped. In this way, two domains complement with each other in the sequence modeling process.

\vspace{-0.4cm}
\subsection{Smoothed dilated convolution}
\label{sec:sd-conv}
Despite the effectiveness of dilated convolution, it may have the so-called ``gridding artifacts"~{\cite{hamaguchi2018effective}}. That is, consecutive units in the output are computed from completely separate sets of units in the input, and the actual receptive field is different. To handle this problem,Wang and Ji $\emph{et.al.}$~{\cite{wang2018smoothed}} proposed the smoothed dilated convolutions (SD-convs) where the interaction was added among adjacent input units before adopting dilation convolution. A separable and shared (SS) convolution is applied before calculating dilated convolution. Here ``separable" means that each channel deals with its kernel filters independently, and ``shared" denotes that the parameters of kernel filters are shared for different input and output channels. In this paper, all the D-convs are replaced by SD-convs to alleviate the artifacts.

\begin{figure}[t]
	\centering
	\centerline{\includegraphics[width=0.92\columnwidth]{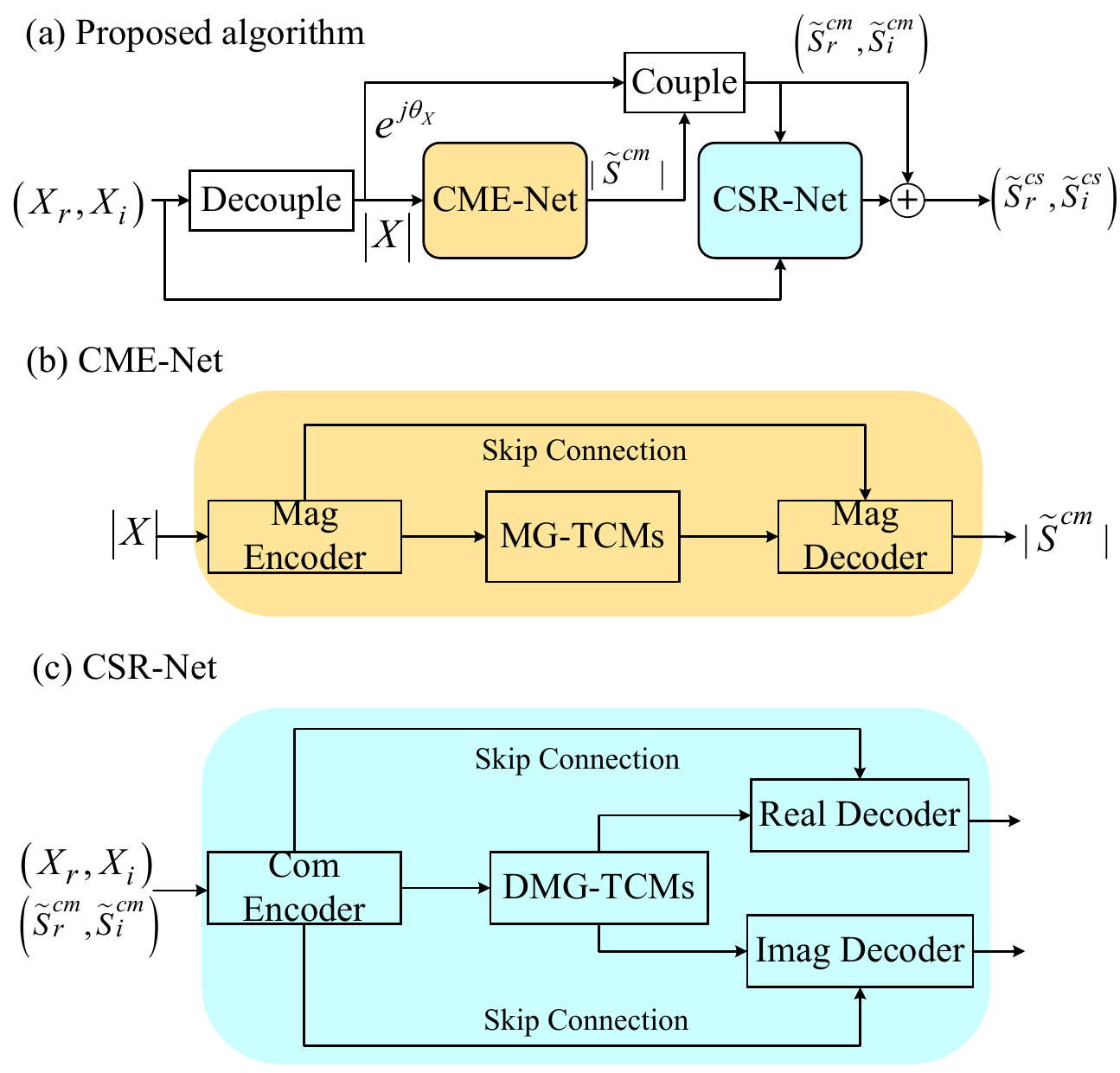}}
	\caption{The diagram of proposed two-stage algorithm. (a) The processing flow of the algorithm. (b) The model topology of CME-Net. (c) The model topology of CSR-Net.}
	\label{fig:two-satge}
	\vspace{-0.6cm}
\end{figure}
\vspace{-0.4cm}
\subsection{Two-stage approach}
\label{sec:two-stage}
Our proposed two-stage algorithm is illustrated in Fig.~{\ref{fig:two-satge}}. It includes two processing stages. In the first stage, CME-Net receives the magnitude of noisy spectrum to coarsely estimate the magnitude of the clean speech, which is then coupled with noisy phase to obtain a coarse complex spectrum. In the second stage, the real and imaginary components for both coarse and noisy spectra are concatenated as the inputs of CSR-Net. Instead of directly refining the complex spectrum, it only captures the spectral details which might be lost in the first stage. Formally, the calculation process is formulated as:
\vspace{-0.2cm}
\begin{gather}
\label{eqn:equa3}
\lvert \tilde{S}^{cm} \rvert = \mathcal{G}_{cm} \left( \lvert X \rvert; \phi_{cm} \right),\\
\tilde{S}^{cm}_{r} = \Re \left(\lvert S^{cm} \rvert e^{j\theta_{X}} \right),
\tilde{S}^{cm}_{i} = \Im \left(\lvert S^{cm} \rvert e^{j\theta_{X}} \right),\\
\left( \tilde{S}^{cs}_{r}, \tilde{S}^{cs}_{i} \right) = \left( \tilde{S}^{cm}_{r}, \tilde{S}^{cm}_{i} \right) + \mathcal{G}_{cs} \left(\tilde{S}^{cm}_{r}, \tilde{S}^{cm}_{i}, X_{r}, X_{i}; \phi_{cs} \right),
\end{gather}
where $\mathcal{G}_{cm}$ and $\mathcal{G}_{cs}$ denote the mapping functions for CME-Net and CSR-Net, respectively. $\phi_{cm}$ and $\phi_{cs}$ denote the parameter sets for the first and second network, respectively. $\tilde S^{cm}$ and $\tilde S^{cs}$ are the estimated outputs from CME-Net and CSR-Net, respectively. $\Re$ and $\Im$ refer to real and imaginary operations, respectively.


The network details are shown in Fig.~{\ref{fig:two-satge}}(b)-(c). The overall topology is similar to that of~{\cite{tan2019learning}}, which includes three major components, namely convolutional encoder, decoder, and sequence modeling module. Instead of using LSTMs as the sequence module, we adopt TCMs for better sequence learning. For CME-Net, we stack 18 MG-TCMs while 12 DMG-TCMS are adopted in CSR-Net~{\footnote{In this paper, 6 TCMs form a large group, where the dialtion rate in each group is (1, 2, 4, 8, 16, 32)}. This is because DMG-TCM is more powerful to capture both long and short-term temporal information than MG-TCM. In both encoders and decodes, 5 convolutional blocks are adopted, each of which includes one (de)convolution layer, instance normalization~{\cite{ulyanov2016instance}}, and Parametric ReLU (PReLU)~{\cite{he2015delving}}. The kernel size for each (de)convolution layers is (2, 3) except (2, 5) for the first layer in the time and frequency axis, respectively. The stride is set to (1, 2) in the time and frequency axis. The number of channels for each intermediate layer is 64. In the first network, Softplus~{\cite{glorot2011deep}} is used as the output activation function to obtain the magnitude of the spectrum and linear function is used to obtain the RI in the second stage. Note that similar to~{\cite{tan2019learning}}, two decoders are used in CSR-Net to obtain RI estimation.
\vspace{-0.2cm}
\subsection{Loss function}
\label{loss-function}
We take the following strategy to train the network. Firstly, we separately train CME-Net until convergence, and the loss is defined as:
\begin{gather}
\label{eqn:equa4}
\mathcal{L}_{cm} =\left \| \lvert \tilde S^{cm} \rvert - \lvert S \rvert \right \|^{2}_{F},
\end{gather}
Afterward, both the first and the second stages are jointly trained, where the first stage is initialized with the pretrained model, and the overall loss is defined as:
\begin{gather}
\label{eqn:equa5}
\mathcal{L} = \mathcal{L}_{cs}^{RI} + \mathcal{L}_{cs}^{Mag} + \lambda \mathcal{L}_{cm},\\
\mathcal{L}_{cs}^{RI} = \left \| \tilde{S}^{cs}_{r} - S_{r} \right \|^{2}_{F} + \left \| \tilde{S}^{cs}_{i} - S_{i} \right \|^{2}_{F},\\
\mathcal{L}_{cs}^{Mag} = \left \| \sqrt{ \lvert \tilde{S}_{r}^{cs}\rvert ^2 + \lvert \tilde{S}_{i}^{cs} \rvert ^2  } - \sqrt{ \lvert S_{r}\rvert ^2 + \lvert S_{i} \rvert ^2} \right \|_{F}^{2},
\end{gather}
where $\mathcal{L}^{RI}$ and $\mathcal{L}^{Mag}$ denote the loss optimization toward RI and that toward magnitude, respectively. $\lambda \in [0,1]$ controls the loss weight of the first network. In this paper, $\lambda$ is set to 0.1.

\vspace{-0.2cm}
\section{EXPERIMENTAL RESULTS}
\label{sec:experiment-result}
\vspace{-0.2cm}

\subsection{Datasets}
We use the WSJ0-SI84 dataset~{\cite{paul1992design}} for evaluation, which includes 7138 utterances by 83 speakers (42 males and 41 females). 5428, and 957 utterances with 77 speakers are split for training and validation, respectively. Two types are set for test. For the first type, the speaker information is within the training dataset (dubbed seen speaker), while for the second type, the speaker information is untrained (dubbed unseen speaker). Each of the types includes 150 utterances. We randomly select 20,000 noises from the DNS-Challenge~{\footnote{https://github.com/microsoft/DNS-Challenge}} to obtain a 55 hours noise set for training. During each mixed process, a random cut is generated to obtain a noise vector, which is subsequently mixed with randomly selected clean utterance. The SNR range for training is [-5\rm{dB}, 0\rm{dB}] with the interval 1\rm{dB}. As a result, totally 50,000, 4000 noisy-clean pairs are established for training and validation, respectively. The total duration for training set is about 100 hours.

For test, we select two noises from NOISEX92~{\cite{varga1993assessment}}, namely babble and factory1. Three SNRs are used for model evaluation, namely -5\rm{dB}, 0\rm{dB} and 5\rm{dB}.

\renewcommand\arraystretch{1.62}
\begin{table*}[t]
	\setcounter{table}{1}
	\caption{Objective result comparisons among different models in terms of PESQ, ESTOI and SDR for both seen and unseen speaker cases. \textbf{BOLD} indicats the best score in each case.}
	\Huge
	\centering
	\resizebox{0.82\textwidth}{!}{
		\begin{tabular}{cc|c|cccccccc|cccccccc|cccccccc}
			\hline
			 &\textbf{Metrics} &\multirow{3}*{\rotatebox{90}{\textbf{Causality}}} &\multicolumn{8}{c|}{\textbf{PESQ}} &\multicolumn{8}{c|}{\textbf{ESTOI(\%)}} &\multicolumn{8}{c}{\textbf{SDR(dB)}} \\
			\cline{1-2}\cline{4-27}
			& \textbf{Noises} &  &\multicolumn{4}{c|}{\textbf{Babble}} &\multicolumn{4}{c|}{\textbf{Factory1}} &\multicolumn{4}{c|}{\textbf{Babble}} &\multicolumn{4}{c|}{\textbf{Factory1}} &\multicolumn{4}{c|}{\textbf{Babble}} &\multicolumn{4}{c}{\textbf{Factory1}} \\
			\cline{1-2}\cline{4-27}
			& \textbf{SNR(dB)} & &-5 &0 &5 &\multicolumn{1}{c|}{Avg.} &-5 &0 &5 &Avg. &-5 &0 &5 &\multicolumn{1}{c|}{Avg.} &-5 &0 &5 &\multicolumn{1}{c|}{Avg.} &-5 &0 &5 &\multicolumn{1}{c|}{Avg.} &-5 &0 &5 &Avg.\\
			\cline{1-27}
			\multirow{6}*{\rotatebox{90}{\textbf{Seen speaker}}}
			&\multicolumn{1}{|c|}{\textbf{Noisy}} &-  &1.64 &1.88 &2.19 &\multicolumn{1}{c|}{1.90} &1.51 &1.79 &2.13 &1.81 &29.32 &42.96 &58.74 &\multicolumn{1}{c|}{43.67} &28.89 &44.41 &60.73 &44.68 &-4.91 &0.04 &5.03 &\multicolumn{1}{c|}{0.05} &-4.90 &0.04 &5.03 &0.06\\
			\cline{2-27}
			&\multicolumn{1}{|c|}{\textbf{CRN}} &$\checkmark$ &1.86 &2.34 &2.76 &\multicolumn{1}{c|}{2.32} &2.01 &2.45 &2.84 &2.43 &49.10 &65.38 &77.35 &\multicolumn{1}{c|}{63.94} &48.63 &65.99 &77.87 &64.16 &1.55 &6.28 &10.45 &\multicolumn{1}{c|}{6.09} &3.36 &7.51 &11.18 &7.35\\
			\cline{2-27}
			&\multicolumn{1}{|c|}{\textbf{TCNN}} &$\checkmark$  &1.91 &2.42 &2.83 &\multicolumn{1}{c|}{2.39} &2.06 &2.50 &2.84 &2.47 &55.12 &70.91 &81.13 &\multicolumn{1}{c|}{69.05} &54.70 &71.04 &80.65 &68.80 &4.55 &9.33 &12.98 &\multicolumn{1}{c|}{8.95} &6.07 &10.07 &13.15 &9.76\\
			\cline{2-27}
			&\multicolumn{1}{|c|}{\textbf{GCRN}} &$\checkmark$  &2.02 &2.53 &2.91 &\multicolumn{1}{c|}{2.49} &2.14 &2.63 &3.00 &2.59 &54.18 &70.07 &80.37 &\multicolumn{1}{c|}{68.21} &54.14 &71.41 &81.12 &68.89 &3.96 &8.44 &11.92 &\multicolumn{1}{c|}{8.11} &6.00 &9.57 &12.66 &9.41\\
			\cline{2-27}
			&\multicolumn{1}{|c|}{\textbf{CME-Net(Pro.)}} &$\checkmark$ &1.98 &2.47 &2.86 &\multicolumn{1}{c|}{2.44} &2.12 &2.55 &2.91 &2.53 &51.77 &67.46 &78.72 &\multicolumn{1}{c|}{65.99} &51.44 &67.69 &78.76 &65.96 &2.32 &6.85 &10.87 &\multicolumn{1}{c|}{6.68} &3.76 &7.76 &11.47 &7.66\\
			\cline{2-27}
			&\multicolumn{1}{|c|}{\textbf{CTS-Net(Pro.)}} &$\checkmark$  &\textbf{2.18} &\textbf{2.75} &\textbf{3.13} &\multicolumn{1}{c|}{\textbf{2.69}} &\textbf{2.31} &\textbf{2.81} &\textbf{3.14} &\textbf{2.76} &\textbf{60.10} &\textbf{75.37} &\textbf{84.12} &\multicolumn{1}{c|}{\textbf{73.19}} &\textbf{59.82} &\textbf{75.05} &\textbf{83.37} &\textbf{72.74} &\textbf{5.46} &\textbf{10.21} &\textbf{13.63} &\multicolumn{1}{c|}{\textbf{9.77}} &\textbf{6.93} &\textbf{10.74} &\textbf{13.70} &\textbf{10.46}\\
			\hline
			\multirow{6}*{\rotatebox{90}{\textbf{Unseen speaker}}}
			&\multicolumn{1}{|c|}{\textbf{Noisy}} &$\checkmark$ &1.55 &1.83 &2.15 &\multicolumn{1}{c|}{1.84} &1.43 &1.74 &2.10 &1.76 &26.60 &39.60 &54.43 &\multicolumn{1}{c|}{40.21} &26.66 &41.01 &57.03 &41.57 &-4.91 &-0.05 &5.03 &\multicolumn{1}{c|}{0.02} &-4.93 &0.04 &5.03 &0.05\\
			\cline{2-27}
			&\multicolumn{1}{|c|}{\textbf{CRN}} &$\checkmark$ &1.76 &2.28 &2.70 &\multicolumn{1}{c|}{2.24} &1.95 &2.40 &2.78 &2.38 &45.31 &62.18 &75.12 &\multicolumn{1}{c|}{60.87} &46.03 &63.07 &75.95 &61.68 &1.43 &6.37 &10.63 &\multicolumn{1}{c|}{6.14} &3.37 &7.63 &11.45 &7.48\\
			\cline{2-27}
			&\multicolumn{1}{|c|}{\textbf{TCNN}} &$\checkmark$  &1.83 &2.36 &2.78 &\multicolumn{1}{c|}{2.32} &2.00 &2.45 &2.80 &2.41 &51.88 &68.92 &79.86 &\multicolumn{1}{c|}{66.89} &51.59 &68.95 &79.55 &66.70 &4.33 &9.35 &13.10 &\multicolumn{1}{c|}{8.93} &6.16 &10.13 &13.39 &9.89\\
			\cline{2-27}
			&\multicolumn{1}{|c|}{\textbf{GCRN}} &$\checkmark$ &1.94 &2.49 &2.90 &\multicolumn{1}{c|}{2.44} &2.13 &2.61 &2.98 &2.57 &52.19 &69.09 &79.88 &\multicolumn{1}{c|}{67.05} &52.85 &70.12 &80.30 &67.76 &3.97 &8.73 &12.37 &\multicolumn{1}{c|}{8.36} &5.85 &9.81 &13.03 &9.56\\
			\cline{2-27}
			&\multicolumn{1}{|c|}{\textbf{CME-Net(Pro.)}} &$\checkmark$  &1.91 &2.42 &2.81 &\multicolumn{1}{c|}{2.38} &2.08 &2.49 &2.86 &2.48 &48.53 &65.20 &76.84 &\multicolumn{1}{c|}{63.53} &49.07 &65.22 &77.23 &63.84 &2.47 &7.22 &11.19 &\multicolumn{1}{c|}{6.96} &4.04 &7.97 &11.81 &7.94\\
			\cline{2-27}
			&\multicolumn{1}{|c|}{\textbf{CTS-Net(Pro.)}} &$\checkmark$ &\textbf{2.11} &\textbf{2.72} &\textbf{3.11} &\multicolumn{1}{c|}{\textbf{2.65}} &\textbf{2.29} &\textbf{2.77} &\textbf{3.12} &\textbf{2.72} &\textbf{57.82} &\textbf{74.14} &\textbf{83.15} &\multicolumn{1}{c|}{\textbf{71.70}} &\textbf{58.07} &\textbf{73.59} &\textbf{82.70} &\textbf{71.45} &\textbf{5.40} &\textbf{10.46} &\textbf{13.96} &\multicolumn{1}{c|}{\textbf{9.94}} &\textbf{7.24} &\textbf{10.96} &\textbf{14.14} &\textbf{10.78}\\
			\hline
	\end{tabular}}
	\label{tbl:objective-results}
	\vspace{-0.4cm}
\end{table*}
\vspace{-0.2cm}
\subsection{Parameter setup}
\label{parameter-setup}
All the utterances are sampled at 16kHz. The 20ms Hanning window is utilized, with 50\% overlap between adjacent frames. 320 point FFT is used. Both models are optimized by Adam~{\cite{kingma2014adam}} with $\beta_{1} = 0.9, \beta_{2} = 0.999$. In the first stage, the learning rate (LR) is set to 0.001. In the second stage, the pre-trained model in the first stage is fine-tuned with LR = 0.0001, while 0.001 for the second model. The batch size is set to 16 at an utterance level, where the maximum utterance length is chunked to 8 seconds for training stability.
\vspace{-0.2cm}
\subsection{Baselines}
We adopt three state-of-the-art baselines for comparison, namely CRN~{\cite{tan2018convolutional}}, TCNN~{\cite{pandey2019tcnn}} and GCRN~{\cite{tan2019learning}}. CRN is a typical convolutional recurrent network with encoder-decoder architecture, and only magnitude is estimated. GCRN is an advanced complex spectral mapping network based on CRN, where both magnitude and phase are estimated. Note that both RI and magnitude are optimized with $\mathcal{L}^{RI}$ and $\mathcal{L}^{Mag}$ for GCRN. For TCNN, the waveform is directly used as both input and target, and stacked TCMs are adopted for sequence modeling. All the models are trained with causal configurations, $\emph{i.e.}$, no future information is involved. The sound demo samples are available online~{\footnote{https://github.com/Andong-Li-speech/CTS-Net}}.

\vspace{-0.2cm}

\subsection{Results and analysis}
\label{results-analysis}
\vspace{-0.2cm}
In this study, PESQ~{\cite{rix2001perceptual}}, ESTOI~{\cite{jensen2016algorithm}} and SDR~{\cite{vincent2007first}} are chosen as objective measurement metrics.
\vspace{-0.0cm}

\renewcommand\arraystretch{0.6}
\begin{table}[t]
	\setcounter{table}{0}
	\caption{Ablation study $\emph{w.r.t.}$ TCM type and SD-conv. TCM(I) and TCM(II) denote the TCMs used in the first and second stage, respectively. ``O", ``MG" and ``DMG" denote O-TCM, MG-TCM and DMG-TCM. All the values are averaged on the test set.}
	\large
	\centering
	\resizebox{0.40\textwidth}{!}{
		\begin{tabular}{ccccccc}
			\toprule
			Models
			& TCM(I)  &TCM(II) &SD  &PESQ &ESTOI(\%) &SDR (dB)\\
			\midrule
			Noisy &- &- &- &1.83 &42.53 &0.30\\
			\midrule
			CME-Net  &O &- &$\times$  &2.40 &64.03 &7.38 \\
			\midrule
			CME-Net &MG &- &$\times$ &2.44 &64.53 &7.35\\
			\midrule
			CTS-Net &MG &MG &$\checkmark$ &2.70 &71.89 &10.16 \\
			\midrule
			CTS-Net &MG &DMG &$\checkmark$ &$\textbf{2.71}$ &$\textbf{72.27}$ &$\textbf{10.24}$\\
			\midrule
			CTS-Net &MG &DMG &$\times$ &2.68 &71.75 &10.11\\
			\bottomrule
		\end{tabular}}
	\label{tbl:ablation-study}
\vspace{-0.4cm}
\end{table}
\vspace{-0.4cm}

\subsubsection{Ablation study}
\label{ablation-study}
\vspace{-0.0cm}
We investigate the effects of different TCMs and SD-conv, as shown in Table~{\ref{tbl:ablation-study}}. From the results, we can have the following observations. Firstly, MG-TCM achieves overall better performance than O-TCM. For example, when only the first stage is trained, CME+MG-TCM obtains 0.04 and 0.50\% improvements in PESQ and ESTOI than CME+O-TCM, which validates the parameter redundancy for original TCM. Secondly, when MG-TCM is replaced by its dual version in the second stage, consistently better performance is achieved in all three metrics. Thirdly, compared with regular dilated convolution, SD-conv provides 0.03, 0.52\% and 0.13\rm{dB} improvements in PESQ, ESTOI and SDR, respectively.
\vspace{-0.0cm}
\subsubsection{Comparison with baselines}
\label{comparison-with-baseline}
\vspace{-0.2cm}
The results of different models are shown in Table~{\ref{tbl:objective-results}}. Note that CME-Net and CTS-Net are the best configurations from the ablation study. One can have the following observations. Firstly, CME-Net significantly outperforms CRN in different cases. For example, for seen speaker, CME-Net provides average 0.12 and 0.10 PESQ improvements than CRN on babble and factory1 noises, while 2.05\% and 1.80\% improvements in ESTOI. This indicates the superior performance of MG-TCMs at sequence modeling capability than naive LSTMs. Secondly, compared with CME-Net, when the second stage is leveraged to refine the spectrum, large metric improvements are achieved. For example, going from CME-Net to CTS-Net, average 0.25 and 6.94\% improvements are achieved in PESQ and ESTOI, respectively. This reveals the necessity and significance of the second stage in improving the speech quality and intelligibility. Thirdly, the proposed two-stage model consistently surpasses all the baselines. For example, compared with GCRN, average 0.18, 4.29\%, and 1.38\rm{dB} metric improvements are obtained in terms of PESQ, ESTOI and SDR, respectively.

Table~{\ref{tbl:model-parameters}} summarizes the number of trainable parameters among different models. One can find that the CME-Net has the smallest number of parameters among different models. As CTS-Net includes two cascaded sub-networks, the number of parameters is relatively large. Nevertheless, it is still less than another three baselines.

\renewcommand\arraystretch{1.0}
\begin{table}[t]
	\setcounter{table}{2}
	\caption{The number of trainable parameters among different models. The unit is million. $\textbf{BOLD}$ denotes the lowest trainable parameters. }
	\centering
	\footnotesize
	\resizebox{0.45\textwidth}{!}{
	\begin{tabular}{c|c|c|c|c|c}
		\hline
		Model &CRN &GCRN &TCNN &CME-Net &CTS-Net\\
		\hline
		Para. (million) &17.59 &9.06 &5.08 &\textbf{1.96} &4.99 \\
		\hline
	\end{tabular}}
	\label{tbl:model-parameters}
	\vspace*{-\baselineskip}
	\vspace{-0.1cm}
\end{table}

\vspace{-0.2cm}
\section{CONCLUSIONS}
\label{sec:conclusion}
\vspace{-0.2cm}
In this work, we propose a two-stage algorithm for monaural noise reduction in the complex domain. In the first stage, the magnitude of spectrum is estimated, which is coupled with noisy phase to obtain a coarse complex spectrum. In the second stage, the spectral details are captured, which further refines the magnitude and phase information simultaneously. Additionally, a modified TCM is proposed, which can achieve better performance than previous counterparts in sequence learning while decreasing the parameter redundancy by a large margin. Experimental results indicate that the proposed algorithm consistently outperforms previous powerful methods while still enjoying a small parameter capacity.
\vspace{-0.2cm}
\section{ACKNOWLEDGMENT}
\label{sec:acknowledge}
\vspace{-0.2cm}
This work was supported by National Key R\&D Program of China. This work was also supported by IACAS Young Elite Researcher Project under no.QNYC201813. We would like to thank Cunhang Fan at Institute of Automation, and Shan You at SenseTime Research for constructive comments.



\vfill\pagebreak

\bibliographystyle{IEEEbib}
\bibliography{refs}

\end{document}